\documentclass[twocolumn]{aastex631}

\definecolor{MyGreen}{rgb}{0.0,0.6,0.3}
\definecolor{MyBlue}{rgb}{0.0,0.3,0.6}
\hypersetup{colorlinks=true,citecolor=MyBlue,linkcolor=MyGreen}
\usepackage{etoolbox, color, soul}
\usepackage{fancyvrb}
\usepackage{float}
\makeatletter 
 \patchcmd{\NAT@citex}
   {\@citea\NAT@hyper@{%
     \NAT@nmfmt{\NAT@nm}%
     \hyper@natlinkbreak{\NAT@aysep\NAT@spacechar}{\@citeb\@extra@b@citeb}%
     \NAT@date}}
   {\@citea\NAT@nmfmt{\NAT@nm}%
   \NAT@aysep\NAT@spacechar\NAT@hyper@{\NAT@date}}{}{}

 \patchcmd{\NAT@citex}
   {\@citea\NAT@hyper@{%
     \NAT@nmfmt{\NAT@nm}%
     \hyper@natlinkbreak{\NAT@spacechar\NAT@@open\if*#1*\else#1\NAT@spacechar\fi}%
       {\@citeb\@extra@b@citeb}%
     \NAT@date}}
   {\@citea\NAT@nmfmt{\NAT@nm}%
   \NAT@spacechar\NAT@@open\if*#1*\else#1\NAT@spacechar\fi\NAT@hyper@{\NAT@date}}
   {}{}
\makeatother

\newcommand{\code}[1]{\texttt{#1}}

\newcommand{\Athena}{\code{Athena++}}

\newcommand{\hyphen}{\,--\,}
\newcommand{\Prad}{P_\mathrm{r}}
\newcommand{\kmps}{$\,$km$\,$s$^{-1}\,$}

\newtoggle{referee}            %
\togglefalse{referee}          %

\iftoggle{referee}{
  \usepackage{color,soul}  
 
  \soulregister\citet7
  \soulregister\citep7
  \soulregister\ref7
}{
}                              %

\usepackage{xcolor}
\usepackage{amsmath}
\usepackage{CJK}

\submitjournal{ApJL}

\shorttitle{SLF Variability in 3D RHD Models}
\shortauthors{Schultz, Bildsten, and Jiang}

\graphicspath{{./}{figures/}}

\begin{document}
\begin{CJK*}{UTF8}{gbsn}

\title{Stochastic Low Frequency Variability in 3-Dimensional Radiation Hydrodynamical Models of Massive Star Envelopes \footnote{Released on ??, ??, 202?}}

\author[0000-0002-0786-7307]{William C. Schultz}
\affiliation{Department of Physics, University of California, Santa Barbara, CA 93106, USA}

\author{Lars Bildsten}
\affiliation{Department of Physics, University of California, Santa Barbara, CA 93106, USA}
\affiliation{Kavli Institute for Theoretical Physics, University of California, Santa Barbara, CA 93106, USA}

\author[0000-0002-2624-3399]{Yan-Fei Jiang(姜燕飞)}
\affiliation{Center for Computational Astrophysics, Flatiron Institute, New York, NY 10010, USA}

\begin{abstract}
Increasing main sequence stellar luminosity with stellar mass leads to the eventual dominance of radiation pressure in stellar envelope hydrostatic balance. 
As the luminosity approaches the Eddington limit, additional instabilities (beyond conventional convection) can occur.
These instabilities readily manifest in the outer envelopes of OB stars, where the opacity increase associated with iron yields density and gas pressure inversions in 1D models.
Additionally, recent photometric surveys (e.g. TESS) have detected excess broadband low frequency variability in power spectra of OB star lightcurves, called stochastic low frequency variability (SLFV).
This motivates our novel 3D Athena++ radiation hydrodynamical (RHD) simulations of two 35$\,$M$_\odot$ star envelopes (the outer $\approx15\%$ of the stellar radial extent), one on the zero-age main sequence and the other in the middle of the main sequence.
Both models exhibit turbulent motion far above and below the conventional iron opacity peak convection zone (FeCZ), obliterating any ``quiet" part of the near-surface region and leading to velocities at the photosphere of 10\hyphen100\kmps, directly agreeing with spectroscopic data.
Surface turbulence also produces SLFV in model lightcurves with amplitudes and power-law slopes that are strikingly similar to those of observed stars.
The characteristic frequencies associated with SLFV in our models are comparable to the thermal time in the FeCZ ($\approx$3\hyphen7$\,$days$^{-1}$).
These simulations, which have no free parameters, are directly validated by observations and, though more models are needed, we remain optimistic that 3D RHD models of main sequence O star envelopes exhibit SLFV originating from the FeCZ.
\end{abstract}


\newpage


\section{Introduction} \label{sec:intro}

The outer envelopes of OB stars present some of the most challenging environments to the current treatment of stellar convection. 
For stars with $M\gtrsim25$\,M$_\odot$, the opacity ($\kappa$) increase associated with iron at $T\approx180,000\,$K causes the luminosity ($L$) to surpass the Eddington value, $L_\mathrm{Edd}=4\pi GMc/\kappa$, which yields density and gas pressure inversions in 1D stellar models \citep{Joss1973,Paxton2013} and a vigorous 3D instability at an optical depth, $\tau_\mathrm{Fe}$.
The instability stymies 1D models when the iron opacity peak is located sufficiently close to the surface to develop significant turbulence.

Convective efficiency associated with the turbulence can be quantified by the parameter $\gamma\sim P_\mathrm{th}v_\mathrm{c}\tau/P_\mathrm{rad}c$, where $v_\mathrm{c}$ is the convective velocity, $c$ is the speed of light, and the total thermal pressure, $P_\mathrm{th}=P_\mathrm{g}+P_\mathrm{rad}$, where $P_\mathrm{g}$ and $P_\mathrm{rad}$ are the gas and radiation pressure \citep{Cox1968}. 
This is essentially comparing convective transport to radiative diffusion. 
When $\gamma<1$, substantial radiation losses occur in moving fluid elements, defining a critical optical depth $\tau_\mathrm{crit}\sim cP_\mathrm{rad}/v_\mathrm{c}P_\mathrm{th}$. 
In massive stars, $\tau_\mathrm{crit}\gtrsim1000$, yielding large regions of these stellar envelopes susceptible to inefficient convection which has only recently been achieved in full 3D radiation hydrodynamical (RHD) simulations \citep{Jiang2015,Jiang2018}.
Additionally, the turbulent convection in the iron opacity peak convection zone (FeCZ) likely plays a dynamical role, exciting plumes that may reach the surface and cause the long-observed micro-turbulence measured in spectral lines of these stars \citep{Cantiello2009}.
Such a mechanism may well be playing a role, but 3D calculations of this regime \citep{Jiang2015,Jiang2017,Jiang2018} found even more surprising properties of these massive star envelopes.
For example, the 3D calculations revealed that the velocity and density fluctuations from the FeCZ propagate well out to the stellar photosphere \citep{Jiang2015,Jiang2018}, eliminating the intervening radiative layer predicted in 1D models.
These simulations also revealed the first understanding of the complex interplay of convective and radiative transport and how it depends on the ratio $\tau_\mathrm{Fe}/\tau_\mathrm{crit}$.
Even further out in the envelope, helium recombination causes an even larger increase in opacity that can lead to continuum driven winds, possibly impacting our understanding of Luminous Blue Variable eruptions \citep{Jiang2018}. 

This complex interplay of convection and radiation is now being probed by photometric observations of these stars from space-based telescopes (e.g. TESS \citep{Ricker2015}) which have detected ubiquitous low amplitude temporal brightness variability. 
Regardless of their spectral class, metallicity ($Z$), or rotation rate, all massive stars exhibit broad-band photometric variability up to $5\,$mmag ($\approx 0.5\%$) on timescales of hours to days \citep{Bowman2019a,Bowman2019b,Bowman2020a,Bowman2020b}, which is referred to as stochastic low frequency variability (SLFV). 
These same stars also exhibit large-scale surface velocity fluctuations with macro-turbulence velocities of 10\hyphen120\kmps that are detected with ground-based, high resolution spectroscopy \citep{Simon-Diaz2017}. 

The origin of SLFV and the large-scale velocity fluctuations is debated.
One possible cause is internal gravity waves (IGWs) generated in the convective hydrogen burning cores that
propagate through the radiative envelope and manifest near the stellar surface \citep{Aerts2009,Aerts2015,Lecoanet2019}.
However, their observability depends on the efficiency of both the excitation processes in the core and the propagation through the envelope. 
Significant theoretical \citep{Goldreich1990,Lecoanet2013} and computational \citep{Aerts2009,Aerts2015,Couston2018} investigations of this phenomenon have occurred. 
Inhomogeneities from stellar winds combined with rotational effects have also been proposed as a possible explanation of SLFV \citep{Moffat2008,David-Uraz2017,Simon-Diaz2018}, and hydrodynamical simulations are currently predicting SLFV signatures arising from line-driven wind instabilities \citep{Krticka2018,Krticka2021}. 
A third possible cause is surface disturbances produced by the FeCZ \citep{Cantiello2009,Jiang2015,Cantiello2019,Lecoanet2019}, which we explore here with 3D RHD models of surface convection regions. 


\begin{figure}[t]
\centering
\includegraphics[width=0.48\textwidth]{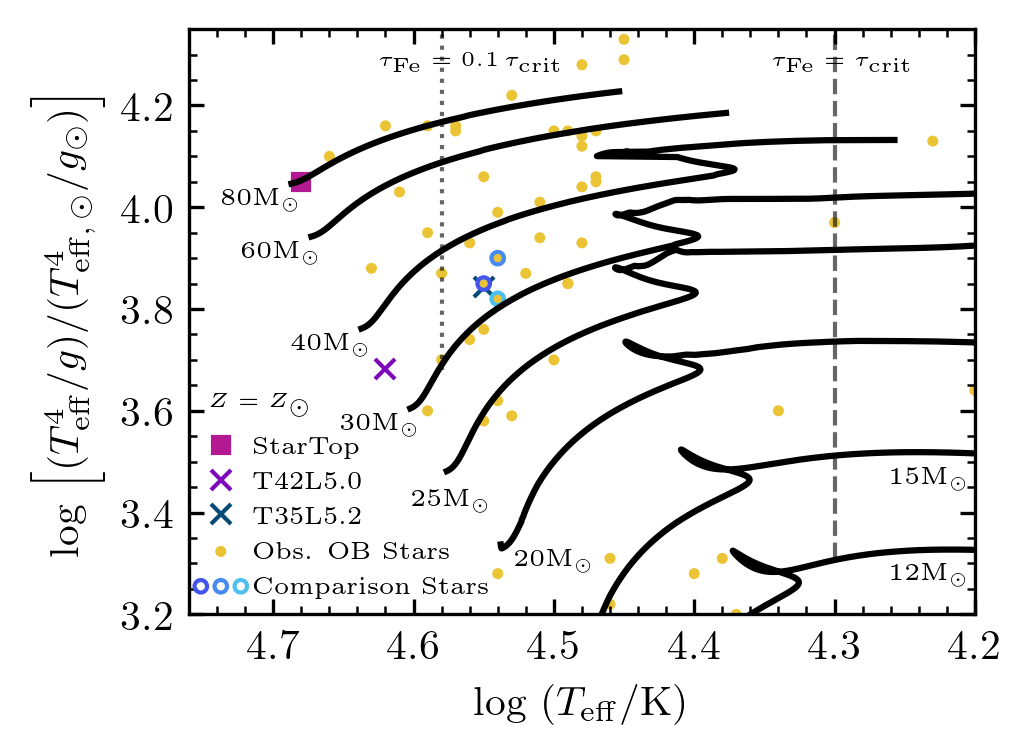}
\caption{\label{fig:HRD} Spectroscopic HR diagram showing locations of new models (the two Xs) relative to recent TESS observations and stellar evolution tracks (the black lines).
The gold circles denote OB star observations from \cite{Burssens2020} while the blue outlines surround stars against which our models are compared in \S~\ref{sec:discussion}. 
The magenta square symbolizes the StarTop model from \cite{Jiang2015}: a plane parallel model of the surface of an 80$\,$M$_{\odot}$ star.
The vertical lines denote locations where the optical depth to the iron opacity peak, $\tau_\mathrm{Fe}$, is equal to (dashed) or 10\% of (dotted) of $\tau_\mathrm{crit}$.}
\end{figure}

\section{3D Models} \label{sec:3D_models}

This work presents two new 3D RHD models: T42L5.0, a Zero-Age Main Sequence (ZAMS) 35$\,$M$_{\odot}$ star, and T35L5.2, a 35$\,$M$_{\odot}$ star half-way through the main sequence.
The model names reflect the effective temperatures, $T_\mathrm{eff}/10^4\,$K, and luminosities, $\log(L/L_{\odot})$.


\begin{figure*}[t]
\centering
\includegraphics[width=0.65\textwidth]{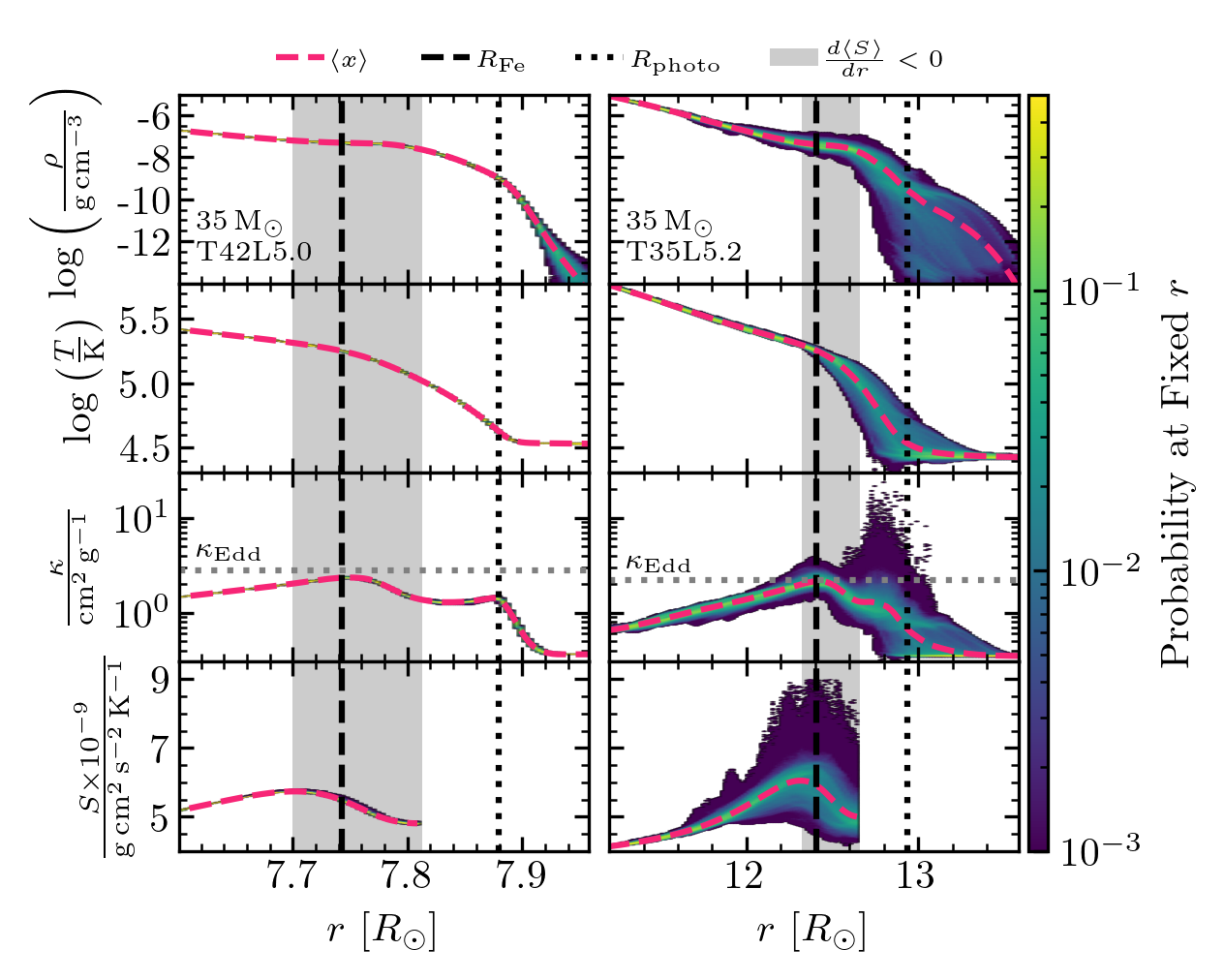}
\caption{\label{fig:T_rho_kap} Profiles of density, temperature, opacity, and entropy (top to bottom respectively) from a single temporal snapshot of T42L5.0 (left) and T35L5.2 (right).
The pink dashed lines show the volume averaged radial profile and the color shows the probability that each quantity has the specified value at that radius.
The sum of the color along vertical lines in each panel is 1.
The vertical dashed black line shows the location of the iron opacity peak and the dotted black line represents the photosphere (where $\langle\tau\rangle=1$).
The gray shaded region denotes where the averaged entropy gradient is negative.
In the $\kappa$ panel, the horizontal line represents the opacity above which the model is super-Eddington.
}
\end{figure*}

\subsection{Computational Methods} \label{sec:comp_methods}
These 3D RHD simulations model massive star envelopes with the code \Athena\ \citep{Stone2020} in spherical polar coordinates. 
The code solves the ideal hydrodynamic equations coupled with the time-dependent, frequency-integrated radiation transport equation for specific intensities over discrete angles based on the numerical algorithm described in \cite{Jiang2021}. 
We use ($128\times 128$) to cover $(\theta,\phi)\in[0.4898\pi,0.5102\pi]\times [0,0.064]$ for model T42L5.0 and $(256\times 256)$ to cover $(\theta,\phi)\in[0.4444\pi,0.5556\pi]\times [0,0.3491]$ for T35L5.2.
Model T42L5.0 utilized 384 logarithmically spaced radial bins to span 6.8$\,$R$_{\odot}$ to 8.2$\,$R$_{\odot}$, while T35L5.2 covers 9.7$\,$R$_{\odot}$ to 15.3$\,$R$_{\odot}$ with 336 logarithmically spaced radial bins. 
Boundary conditions are the same as used in \cite{Jiang2018}.
These simulations take 3000 Skylake cores 4 days to run 1 day of model time.
The gravitational potential is taken to be spherically symmetric, $\phi(r)=-GM/r$, where $G$ is the gravitational constant, $r$ is the radial coordinate and $M$ is the total mass inside $r$.
All the models were run with solar metallicity. 
We calculate opacities using OPAL opacity tables \citep{Iglesias1996} and local densities and temperatures.  
These opacity tables do not include additional line forces in optically thin regions.
Due to computational limitations and the long time required for heat to escape the base of the envelope ($>100\,$days of model time), T42L5.0 and T35L5.2 reached thermal equilibrium down to $r=7.1\,$R$_{\odot}$ and $r=11.1\,$R$_{\odot}$ respectively and both models' FeCZ evolved for nearly 100 thermal times.
The analysis we present in this work only concerns the surface regions of the simulations, in which all properties of interest in the model (e.g. velocities, densities, temperatures) are in thermal equilibrium.

In this work, all 1D comparisons are with reference to Modules for Experiments in Stellar Astrophysics \citep[MESA, ][]{Paxton2011,Paxton2013,Paxton2015,Paxton2018,Paxton2019} models.
Specifically, MESA models generated using the default inlist values from version 15140 were used to both determine the stellar parameters to use as well as the initial conditions for the \Athena\ simulations.
MESA models from \cite{Cantiello2021} were used to show stellar evolutionary tracks in Figure~\ref{fig:HRD}. 


\begin{figure*}[t]
\centering
\includegraphics[width=\textwidth]{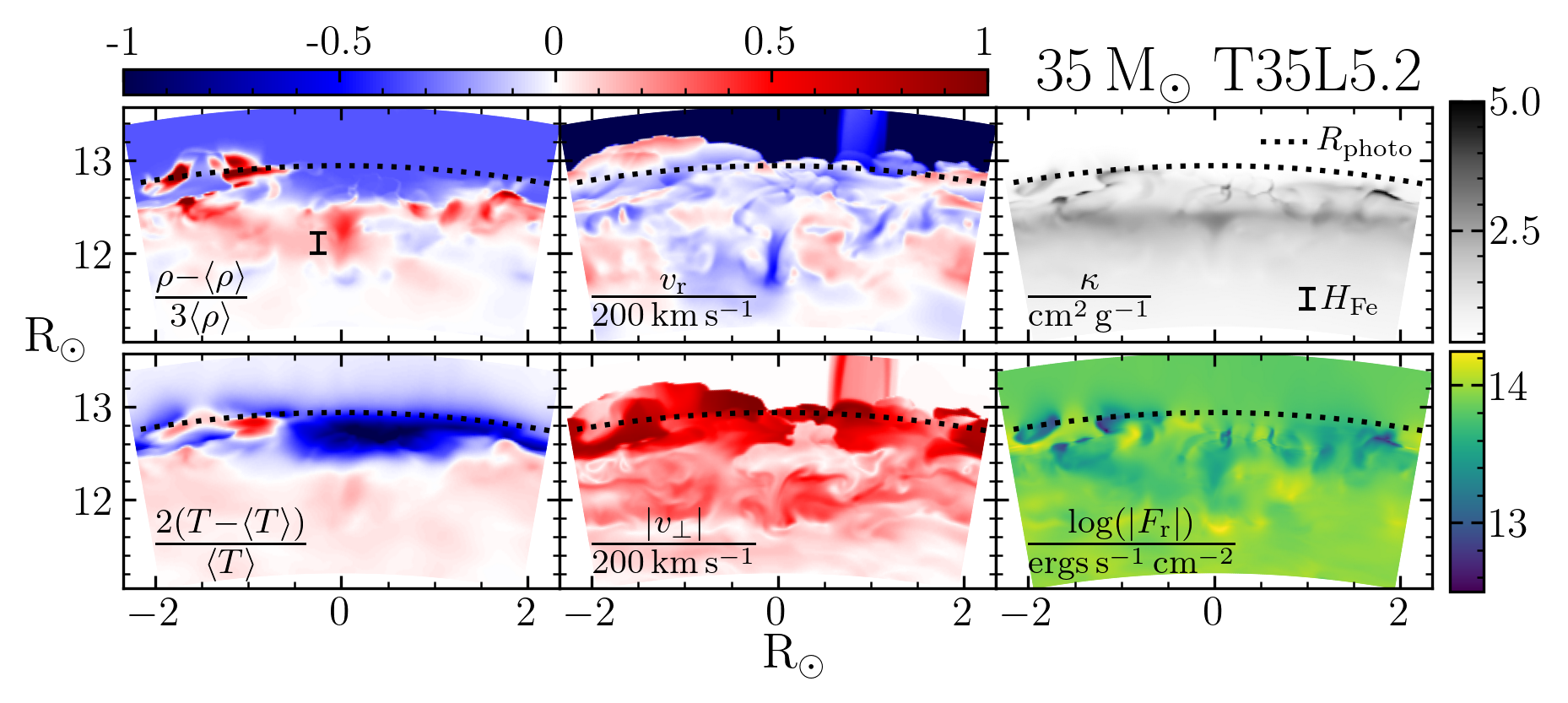}
\caption{\label{fig:slices} Slices of over-density, radial velocity, opacity, diffusive radiative flux, tangential velocity magnitude, and over-temperature (clockwise from upper left respectively) though a $\theta$-r plane of a single temporal snapshot from T35L5.2.
The black dotted line shows the photosphere for the entire snapshot and the scale height at the iron opacity peak, $H_\mathrm{Fe}$, is denoted by the scale-bar.
}
\end{figure*}

\subsection{Model Parameters and Characteristics} \label{sec:model_info}

Figure~\ref{fig:HRD} shows where our 3D models lie in the spectroscopic HR diagram (sHRD) and compares them to previous RHD models, MESA stellar evolution tracks, and recent TESS observations of solar metallicity stars.
T35L5.2 was chosen to closely match three TESS observations and T42L5.0 was chosen to be a more convectively-quiet model for comparison and to probe dynamics in hotter stars with $\tau_\mathrm{crit}\ll\tau_\mathrm{Fe}$.

The stark contrast in variance of fundamental variables is depicted in Figure~\ref{fig:T_rho_kap}, which shows the variations in quantities of interest for single temporal snapshots of both models long after they have reached equilibrium.
T42L5.0 shows nearly no variations throughout the optically thick region.
This is likely because the entire profile of this model is sub-Eddington and convective flux carries $<1\%$ of the total flux.

In contrast, T35L5.2 displays significant variations at each radius including several orders of magnitude of density fluctuations in the optically thick region.
Vigorous convection in the FeCZ carries $\approx12\%$ of the total flux and the motion of the convective plumes cause significant variations at higher altitudes.
As seen in Red Supergiant models \citep{Goldberg2021}, we find that over-densities propagate upwards due to radiative accelerations, but here associated with the He opacity peak near the surface.
Despite the luminosity being larger than the local Eddington luminosity near the Fe and He opacity peaks, the radially-averaged density profile does not have any inversions which plague 1D models.

Shell averaged radial profiles of T35L5.2 are more compact than the initial 1D MESA model due to reduced radiation support caused by correlations between $F_\mathrm{r}$, $\rho$, and $\kappa$ as discussed in \cite{Schultz2020}.
These correlations are visible in Figure~\ref{fig:slices} which displays a $\theta$-r plane slice of a snapshot of T35L5.2. 
This shows that $\rho$ and $\kappa$ are correlated while both being inversely correlated with $F_\mathrm{r}$ leading to a reduced radiative pressure gradient $\nabla\Prad$. 
Not enough turbulence was generated in T42L5.0 for the correlations to lead to substantial reductions in $\nabla\Prad$, however the correlations are still present just outside the photosphere. 

Figure~\ref{fig:slices} also shows the structure of the opacity, density, and both radial and tangential velocity fields, $v_\mathrm{r}$ and $v_{\bot}$ respectively.
The Fe opacity peak can be seen as the gray band-like structure at $r\approx12.7\,R_{\odot}$ and the dark clumps near the photosphere are peaks from the He opacity.
These opaque regions are associated with dense clumps launched from convective plumes, which are as large as the scale height at the iron opacity peak, $H_\mathrm{Fe}$.
The velocity field is significant throughout the near-surface region, with $v_\mathrm{r}$ and $v_{\bot}$ comparable below and at the FeCZ, but $v_{\bot}$ dominates by a factor of a few at and above the photosphere.
As most of the plumes turn around at the photosphere, the radial velocity decreases and the tangential velocity increases slightly as plumes spread out and begin to fall back into the star.


\section{Comparisons to Observations} \label{sec:discussion}

We now compare the surface velocities and photometric variability from the simulations with observations and 1D models. 

\subsection{Near-Surface Convection and Surface Velocities} \label{sec:vel_comp}

\begin{figure}[t]
\centering
\includegraphics[width=0.48\textwidth]{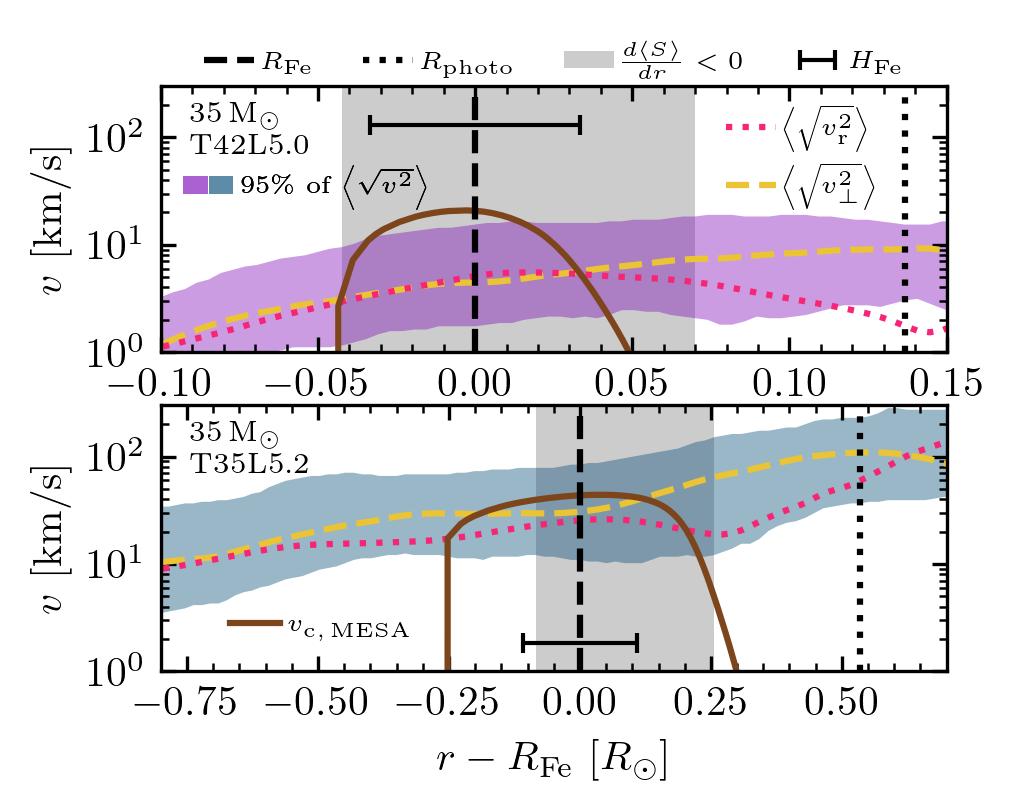}
\caption{\label{fig:vels} Velocity profiles for single temporal snapshots of T42L5.0 (top) and T35L5.2 (bottom) versus radius relative to the location of the iron opacity peak.
The colored shaded regions denote the $95\%$ probability interval of the rms velocities at each radius.
Dashed gold and dotted red lines are the volume weighted averages rms tangential and radial velocities respectively.
The solid brown line denotes the convective velocity profile from analogous 1D MESA models.
The black scale-bar shows the extent of the scale height at the iron opacity peak. 
Vertical black lines and gray shaded regions are the same as Figure~\ref{fig:T_rho_kap}.}
\end{figure}

Figure~\ref{fig:vels} shows the spread of rms velocities throughout our models as well as typical values for the rms tangential and radial velocities, $\langle\sqrt{v_{\bot}^2}\rangle$ and $\langle\sqrt{v_\mathrm{r}^2}\rangle$ respectively.
The rms velocities from the 3D RHD models persist well outside the typically defined FeCZ in 1D models, with both significant undershooting and strong velocity fields at the photosphere.
The velocity profile from the MESA models, shown by the brown line in Figure~\ref{fig:vels}, are confined to the FeCZ though the magnitude of the velocities in these regions are comparable. 

Further, our models predict there is no convectively quiet zone between the FeCZ and the photosphere.
These convective motions, though carrying minimal flux, 
create a turbulent region spanning the outer $\approx7\%$ of the stellar radial extent, propagating to the photosphere where we see typical surface velocities of 9.3\kmps and 123.6\kmps in T42L5.0 and T35L5.2 respectively. 
The three stars similar to T35L5.2 do not have reported macroturbulence velocities, however the full sample of OB stars have $10\leq v_\mathrm{macro}\leq120$\kmps \citep{Burssens2020}.
This agreement is striking considering our models have no free parameters.

Additionally, our models appear to be dominated by tangential velocities at the photosphere, with $\langle\sqrt{v_{\bot}^2}\rangle/\langle\sqrt{v_\mathrm{r}^2}\rangle\sim 2$\hyphen10.
This anisotropy agrees with the recent observational work which determined that radial-tangential fits match observed macroturbulent broadening better than an isotropic Gaussian fit \citep{Simon-Diaz2010,SimonDiaz2014,Simon-Diaz2017} though the extremity of the ratio of the tangential and radial components is debated.
Models of IGW propagation predict drastically more tangential motion compared to radial \citep{Aerts2009}, whereas this work and others predict surface disturbances from the FeCZ produce more isotropic velocity fields \citep{Jiang2015, Jiang2018}.
Determining anisotropies in surface velocities is therefore vital to understanding whether surface velocities are dominated by IGWs or near-surface convection and more models with direct observational comparisons are needed to verify either hypothesis.

With this agreement, our models strongly support the hypothesis that surface velocities, both macro- and micro-turbulence, are affected by near-surface convection regions in OB stars with $M\gtrsim35\,$M$_{\odot}$.


\subsection{Stochastic Low Frequency Variability}

TESS is revealing that OB stars have ubiquitous SLFV. 
Specifically, the $\log{g}$ and $T_\mathrm{eff}$ of T35L5.2 closely match three O stars (HD41997, HD74920, HD326331) observed by TESS (yellow points outlined in blue in Figure~\ref{fig:HRD}).
Unfortunately, no observed stars lie near T42L5.0 in the sHRD so a direct comparison can only be carried out for T35L5.2.

\begin{figure*}[ht]
\centering
\includegraphics[width=0.65\textwidth]{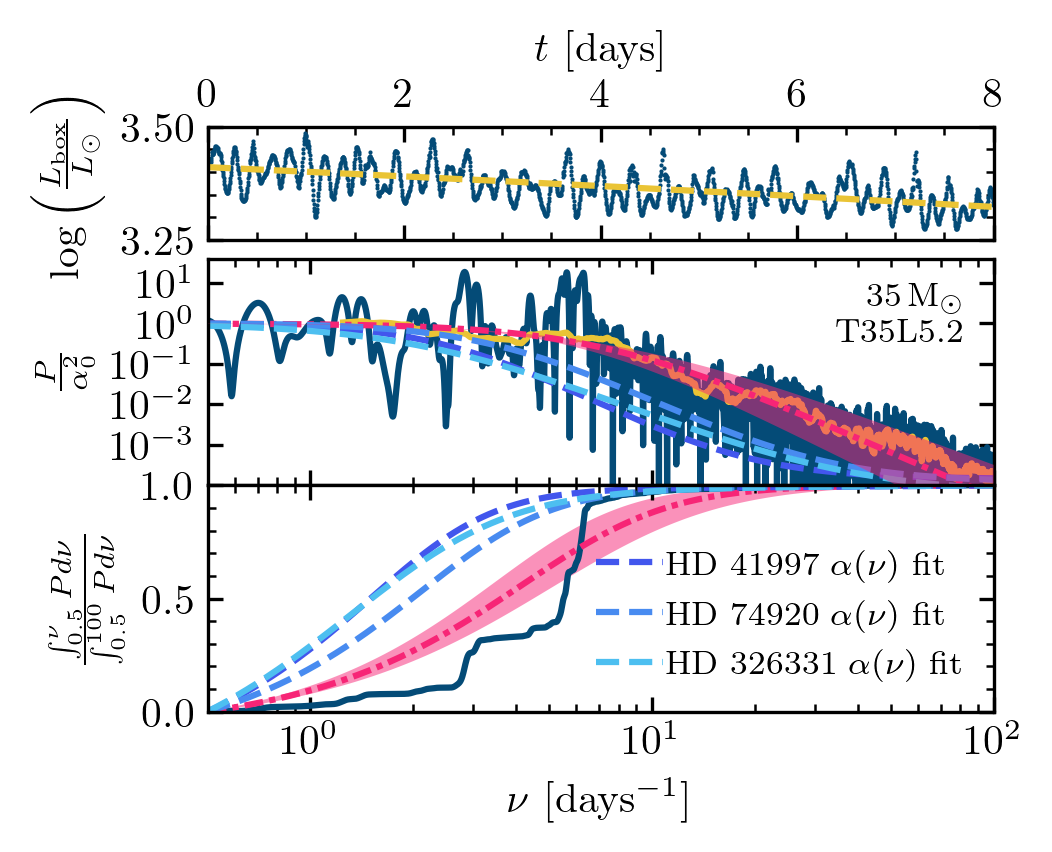}
\caption{\label{fig:SLF} Top: Lightcurve of the T35L5.2 model (dark blue points) for 8 days after the model has reached a steady state equilibrium in the outer regions.
The gold dashed line is a first order polynomial fit of the lightcurve used to zero-mean the lightcurve before taking the power spectrum.
Middle: Power spectrum of the lightcurve from the top panel (solid dark blue line) compared to the normalized SLF fit, $\alpha(\nu)$, from observed OB stars calculated by \cite{Bowman2020a} (dashed blue lines of different shades).
The colors of the dashed lines match the outlines of the observed locations in Figure~\ref{fig:HRD}.
The solid gold line shows the result of removing the periodic signals and linearly smoothing the power spectrum.
The pink dot-dashed line denotes the $\alpha(\nu)$ fit from T35L5.2 and the pink shaded region shows the $95\%$ confidence interval.
Bottom: Cumulative power spectrum of all normalized power spectra on the middle panel normalize to be zero at the left limit ($\nu=0.5\,$days$^{-1}$) and one at the right edge.}
\end{figure*}

Integrated luminosity from model T35L5.2 as a function of time after the FeCZ had safely reached thermal equilibrium is shown in the top panel of Figure~\ref{fig:SLF}.
To quantify the temporal variability, a first order polynomial was fit, subtracted, and divided from the model lightcurve to remove the long timescale decrease in luminosity associated with the lack of thermal equilibrium at the base of the model and calculate $\delta L/L$.
Utilizing the Python code framework \code{SciPy} \citep{Scipy2020}, a Lomb-Scargle periodogram \citep{Lomb1976,Scargle1982,Townsend2010} was calculated from the $\delta L/L$ model lightcurve and the resulting power spectrum, normalized by the low frequency power, $\alpha_0^2$, is shown by the solid blue line in the middle panel of Figure~\ref{fig:SLF}.
Several peaks between $\nu = 2.5$\hyphen7$\,$days$^{-1}$ contain nearly an order of magnitude more power than any other frequency in the power spectrum.
This is further quantified in the bottom panel of Figure~\ref{fig:SLF}, which shows the normalized cumulative power spectrum, as the solid blue line contains 4 cliffs that account for $75\%$ of the total power.
Our analysis of T35L5.2 and T45L5.0 showed these quasi-periodic oscillations are caused by convection driven radial and non-radial pulsations with frequencies and magnitudes that depend on the size of the simulation domain. Properties of these peaks cannot be reliably determined by our simulations.

Aside from the prominent quasi-periodic frequency peaks, SLFV is apparent in T35L5.2. 
Because the peaks are artifacts of our simulation domain, they are removed before a fit to the SLFV is carried out.
Instead of using the sophisticated pre-whitening technique used by \cite{Bowman2019a,Bowman2019b} and others, the peak values were reduced to the mean of the neighboring frequency bins in the power spectrum.
Then a simple moving average was used to smooth out the noise in the power spectrum, the result of which is represented by the solid gold line in the middle panel of Figure~\ref{fig:SLF}.
Following \cite{Bowman2020a}, we used a Bayesian Markov chain Monte Carlo (MCMC) framework with the Python code \code{emcee} \citep{emcee2013} to fit the amplitude of the modified power spectrum to $\alpha(\nu) = \alpha_0 /(1+(\nu/\nu_\mathrm{char})^{\gamma})+C_\mathrm{W}$, where $\nu_\mathrm{char}$ is a characteristic frequency of the knee in the SLFV, $\gamma$ sets the slope of the exponential decay, and $C_\mathrm{W}$ sets the amplitude of the white noise floor.
As our model does not have a white noise floor, $C_\mathrm{W}$ was set to 0 for our fitting.
The resulting fit, shown by the pink dot-dashed line with the 95$\%$ confidence intervals shown by the pink shaded region in Figure~\ref{fig:SLF}, found $\alpha_0=0.0023\pm0.0005$ in $\delta L/L$, $\nu_\mathrm{char}=7.2\pm0.6\,$days$^{-1}$ and $\gamma=1.9\pm0.2$ to be the optimal parameters.
Fits for three comparison stars done by \cite{Bowman2020a} are shown by the dashed blue lines in the middle and bottom panels of Figure~\ref{fig:SLF}.
If we assume the stellar surface is simply a conglomerate of our models arranged in an uncorrelated manner, we would expect $\alpha_0$ to be reduced by up to a factor of $\sqrt{n}$, with $n\approx50$ being the number of models needed to cover the surface, while $\nu_\mathrm{char}$ and $\gamma$ would be unaffected.
Thus we predict $\alpha_0\approx0.033-0.2\%$ variations in $\delta L/L$, agreeing with observed values (0.03-0.1\% variations).
Additionally, the $\gamma$ value of the model's SLFV fit agrees well with the observed values of 1.7\hyphen2.3.
However, the characteristic frequency of our model's power spectral fit, which is consistent with the thermal timescale of the iron opacity peak region ($\approx4\,$hrs), is different than the values observed in similar stars.
Though we have no explanation for this quantitative discrepancy in the exact power spectrum, we remain optimistic that 3D RHD models of main sequence O stars exhibit SLFV originating from the FeCZ.

The power spectrum of the ZAMS model, T42L5.0, is dominated by a quasi-periodic oscillation driven from a fundamental radial pulsation at $\nu\approx 18\,$days$^{-1}$ but still shows the SLFV knee with a smaller amplitude of variability compared to T35L5.2.
However, when using the same technique as described above to fit $\alpha(\nu)$ for T42L5.0 we find our prediction of $\alpha_0\approx3\times10^{-6}$, $\gamma=3.5\pm0.2$, and $\nu_\mathrm{char}=1.2\pm0.05\,$days$^{-1}$ to be potentially consistent with those of similar hot OB star observations (i.e. HD110360 and HD37041).


\section{Conclusion and Future Work} \label{sec:conclusion}

Our two new 3D RHD models show extended FeCZs and contain significant photospheric velocities comparable to those observed in OB stars.
Vigorous, trans-sonic, turbulent convection develops in the middle main sequence model (T35L5.2) causing large variations in $\rho$ and exhibiting a He opacity peak near the surface.
There is no quiet region throughout the outer part of the envelope with surface velocities of $\approx100$\kmps, matching observed macroturbulent velocities as well as showing a slight anisotropy in directionality, with a preference towards tangential versus radial velocities.
Lightcurves from both models show prominent SLFV which agrees in amplitude with observed OB stars, but with more power at higher frequencies than observed.

Inspired by the realization of SLFV in our 3D RHD models, we plan to investigate if stars in other parts of the sHRD with different stellar parameters generate SLFV via near-surface convection zones.
Lower mass stars ($M\sim 10\,$M$_{\odot}$) are substantially less Eddington limited, with weaker turbulent convection near the surface giving rise to a debate about the origin of their SLFV.
The amplitude of the SLFV observed by TESS for these lower mass stars is significantly smaller \citep{Bowman2020a}, which could be explained by the weaker effects of the opacity peaks. 
SLFV is also present in recent observations of the SMC and LMC (e.g. \cite{Kourniotis2014,Bowman2019b,DornWallenstein2020}) and we plan to investigate how metallicity impacts the observed variability with future models. 

Nearly all the OB stars with SLFV have substantial measured surface rotation velocities \citep{Bowman2019a, Bowman2019b, Burssens2020}. 
However, as the inferred rotation periods ($\sim\,$days) are typically much longer than the eddy-turnover times at the opacity peaks ($\sim1\,$hour), we are comfortable with our current exploration neglecting rotational effects. 
Some of the observed OB stars with SLFV are known to have strong magnetic fields (up to $\approx10\,$kG) with recent work highlighting that strong fields can potentially alter surface dynamics \citep{Sundqvist2013,MacDonald2019,Jermyn2020,Cantiello2021}.
Although \Athena\ has the capabilities to include magneto-hydrodynamics \citep[see][]{Jiang2017}, most observed stars with SLFV have fields $<10\,$kG, so we do not see an immediate cause to investigate magnetic field effects. 


\section*{Acknowledgments}
We thank May G. Pedersen and Benny Tsang for many helpful conversations and comments.
This research was supported in part by the NASA ATP grant ATP-80NSSC18K0560, by the National Science Foundation through grant PHY 17-48958 at the KITP.
Resources supporting this work were also provided by the NASA High-End Computing (HEC) programme through the NASA Advanced Supercomputing (NAS) Division at Ames Research Center. 
We acknowledge support from the Center for Scientific Computing from the CNSI, MRL: an NSF MRSEC (DMR-1720256) and NSF CNS-1725797.
The Flatiron Institute is supported by the Simons Foundation.

\bibliography{sample631}{}

\begin{thebibliography}{}
\expandafter\ifx\csname natexlab\endcsname\relax\def\natexlab#1{#1}\fi
\providecommand{\url}[1]{\href{#1}{#1}}
\providecommand{\dodoi}[1]{doi:~\href{http://doi.org/#1}{\nolinkurl{#1}}}
\providecommand{\doeprint}[1]{\href{http://ascl.net/#1}{\nolinkurl{http://ascl.net/#1}}}
\providecommand{\doarXiv}[1]{\href{https://arxiv.org/abs/#1}{\nolinkurl{https://arxiv.org/abs/#1}}}

\bibitem[{{Aerts} {et~al.}(2009){Aerts}, {Puls}, {Godart}, \&
  {Dupret}}]{Aerts2009}
{Aerts}, C., {Puls}, J., {Godart}, M., \& {Dupret}, M.~A. 2009, \aap, 508, 409,
  \dodoi{10.1051/0004-6361/200810471}

\bibitem[{{Aerts} \& {Rogers}(2015)}]{Aerts2015}
{Aerts}, C., \& {Rogers}, T.~M. 2015, \apjl, 806, L33,
  \dodoi{10.1088/2041-8205/806/2/L33}

\bibitem[{{Bowman}(2020)}]{Bowman2020b}
{Bowman}, D.~M. 2020, Frontiers in Astronomy and Space Sciences, 7, 70,
  \dodoi{10.3389/fspas.2020.578584}

\bibitem[{{Bowman} {et~al.}(2020){Bowman}, {Burssens}, {Sim{\'o}n-D{\'\i}az},
  {Edelmann}, {Rogers}, {Horst}, {R{\"o}pke}, \& {Aerts}}]{Bowman2020a}
{Bowman}, D.~M., {Burssens}, S., {Sim{\'o}n-D{\'\i}az}, S., {et~al.} 2020,
  \aap, 640, A36, \dodoi{10.1051/0004-6361/202038224}

\bibitem[{{Bowman} {et~al.}(2019{\natexlab{a}}){Bowman}, {Aerts}, {Johnston},
  {Pedersen}, {Rogers}, {Edelmann}, {Sim{\'o}n-D{\'\i}az}, {Van Reeth},
  {Buysschaert}, {Tkachenko}, \& {Triana}}]{Bowman2019a}
{Bowman}, D.~M., {Aerts}, C., {Johnston}, C., {et~al.} 2019{\natexlab{a}},
  \aap, 621, A135, \dodoi{10.1051/0004-6361/201833662}

\bibitem[{{Bowman} {et~al.}(2019{\natexlab{b}}){Bowman}, {Burssens},
  {Pedersen}, {Johnston}, {Aerts}, {Buysschaert}, {Michielsen}, {Tkachenko},
  {Rogers}, {Edelmann}, {Ratnasingam}, {Sim{\'o}n-D{\'\i}az}, {Castro},
  {Moravveji}, {Pope}, {White}, \& {De Cat}}]{Bowman2019b}
{Bowman}, D.~M., {Burssens}, S., {Pedersen}, M.~G., {et~al.}
  2019{\natexlab{b}}, Nature Astronomy, 3, 760,
  \dodoi{10.1038/s41550-019-0768-1}

\bibitem[{{Burssens} {et~al.}(2020){Burssens}, {Sim{\'o}n-D{\'\i}az}, {Bowman},
  {Holgado}, {Michielsen}, {de Burgos}, {Castro}, {Barb{\'a}}, \&
  {Aerts}}]{Burssens2020}
{Burssens}, S., {Sim{\'o}n-D{\'\i}az}, S., {Bowman}, D.~M., {et~al.} 2020,
  \aap, 639, A81, \dodoi{10.1051/0004-6361/202037700}

\bibitem[{{Cantiello} \& {Braithwaite}(2019)}]{Cantiello2019}
{Cantiello}, M., \& {Braithwaite}, J. 2019, \apj, 883, 106,
  \dodoi{10.3847/1538-4357/ab3924}

\bibitem[{{Cantiello} {et~al.}(2021){Cantiello}, {Lecoanet}, {Jermyn}, \&
  {Grassitelli}}]{Cantiello2021}
{Cantiello}, M., {Lecoanet}, D., {Jermyn}, A.~S., \& {Grassitelli}, L. 2021,
  arXiv e-prints, arXiv:2102.05670.
\newblock \doarXiv{2102.05670}

\bibitem[{{Cantiello} {et~al.}(2009){Cantiello}, {Langer}, {Brott}, {de Koter},
  {Shore}, {Vink}, {Voegler}, {Lennon}, \& {Yoon}}]{Cantiello2009}
{Cantiello}, M., {Langer}, N., {Brott}, I., {et~al.} 2009, \aap, 499, 279,
  \dodoi{10.1051/0004-6361/200911643}

\bibitem[{{Couston} {et~al.}(2018){Couston}, {Lecoanet}, {Favier}, \& {Le
  Bars}}]{Couston2018}
{Couston}, L.-A., {Lecoanet}, D., {Favier}, B., \& {Le Bars}, M. 2018, Journal
  of Fluid Mechanics, 854, R3, \dodoi{10.1017/jfm.2018.669}

\bibitem[{{Cox} \& {Giuli}(1968)}]{Cox1968}
{Cox}, J.~P., \& {Giuli}, R.~T. 1968, {Principles of stellar structure}

\bibitem[{{David-Uraz} {et~al.}(2017){David-Uraz}, {Owocki}, {Wade},
  {Sundqvist}, \& {Kee}}]{David-Uraz2017}
{David-Uraz}, A., {Owocki}, S.~P., {Wade}, G.~A., {Sundqvist}, J.~O., \& {Kee},
  N.~D. 2017, \mnras, 470, 3672, \dodoi{10.1093/mnras/stx1478}

\bibitem[{{Dorn-Wallenstein} {et~al.}(2020){Dorn-Wallenstein}, {Levesque},
  {Neugent}, {Davenport}, {Morris}, \& {Gootkin}}]{DornWallenstein2020}
{Dorn-Wallenstein}, T.~Z., {Levesque}, E.~M., {Neugent}, K.~F., {et~al.} 2020,
  \apj, 902, 24, \dodoi{10.3847/1538-4357/abb318}

\bibitem[{{Foreman-Mackey} {et~al.}(2013){Foreman-Mackey}, {Hogg}, {Lang}, \&
  {Goodman}}]{emcee2013}
{Foreman-Mackey}, D., {Hogg}, D.~W., {Lang}, D., \& {Goodman}, J. 2013, \pasp,
  125, 306, \dodoi{10.1086/670067}

\bibitem[{{Goldberg} {et~al.}(2021){Goldberg}, {Jiang}, \&
  {Bildsten}}]{Goldberg2021}
{Goldberg}, J.~A., {Jiang}, Y.-F., \& {Bildsten}, L. 2021, arXiv e-prints,
  arXiv:2110.03261.
\newblock \doarXiv{2110.03261}

\bibitem[{{Goldreich} \& {Kumar}(1990)}]{Goldreich1990}
{Goldreich}, P., \& {Kumar}, P. 1990, \apj, 363, 694, \dodoi{10.1086/169376}

\bibitem[{{Iglesias} \& {Rogers}(1996)}]{Iglesias1996}
{Iglesias}, C.~A., \& {Rogers}, F.~J. 1996, \apj, 464, 943,
  \dodoi{10.1086/177381}

\bibitem[{{Jermyn} \& {Cantiello}(2020)}]{Jermyn2020}
{Jermyn}, A.~S., \& {Cantiello}, M. 2020, \apj, 900, 113,
  \dodoi{10.3847/1538-4357/ab9e70}

\bibitem[{{Jiang}(2021)}]{Jiang2021}
{Jiang}, Y.-F. 2021, \apjs, 253, 49, \dodoi{10.3847/1538-4365/abe303}

\bibitem[{{Jiang} {et~al.}(2015){Jiang}, {Cantiello}, {Bildsten}, {Quataert},
  \& {Blaes}}]{Jiang2015}
{Jiang}, Y.-F., {Cantiello}, M., {Bildsten}, L., {Quataert}, E., \& {Blaes}, O.
  2015, \apj, 813, 74, \dodoi{10.1088/0004-637X/813/1/74}

\bibitem[{{Jiang} {et~al.}(2017){Jiang}, {Cantiello}, {Bildsten}, {Quataert},
  \& {Blaes}}]{Jiang2017}
---. 2017, \apj, 843, 68, \dodoi{10.3847/1538-4357/aa77b0}

\bibitem[{{Jiang} {et~al.}(2018){Jiang}, {Cantiello}, {Bildsten}, {Quataert},
  {Blaes}, \& {Stone}}]{Jiang2018}
{Jiang}, Y.-F., {Cantiello}, M., {Bildsten}, L., {et~al.} 2018, \nat, 561, 498,
  \dodoi{10.1038/s41586-018-0525-0}

\bibitem[{{Joss} {et~al.}(1973){Joss}, {Salpeter}, \& {Ostriker}}]{Joss1973}
{Joss}, P.~C., {Salpeter}, E.~E., \& {Ostriker}, J.~P. 1973, \apj, 181, 429,
  \dodoi{10.1086/152060}

\bibitem[{{Kourniotis} {et~al.}(2014){Kourniotis}, {Bonanos}, {Soszy{\'n}ski},
  {Poleski}, {Krikelis}, {Udalski}, {Szyma{\'n}ski}, {Kubiak},
  {Pietrzy{\'n}ski}, {Wyrzykowski}, {Ulaczyk}, {Koz{\l}owski}, \&
  {Pietrukowicz}}]{Kourniotis2014}
{Kourniotis}, M., {Bonanos}, A.~Z., {Soszy{\'n}ski}, I., {et~al.} 2014, \aap,
  562, A125, \dodoi{10.1051/0004-6361/201322856}

\bibitem[{{Krti{\v{c}}ka} \& {Feldmeier}(2018)}]{Krticka2018}
{Krti{\v{c}}ka}, J., \& {Feldmeier}, A. 2018, \aap, 617, A121,
  \dodoi{10.1051/0004-6361/201731614}

\bibitem[{{Krti{\v{c}}ka} \& {Feldmeier}(2021)}]{Krticka2021}
---. 2021, \aap, 648, A79, \dodoi{10.1051/0004-6361/202040148}

\bibitem[{{Lecoanet} \& {Quataert}(2013)}]{Lecoanet2013}
{Lecoanet}, D., \& {Quataert}, E. 2013, \mnras, 430, 2363,
  \dodoi{10.1093/mnras/stt055}

\bibitem[{{Lecoanet} {et~al.}(2019){Lecoanet}, {Cantiello}, {Quataert},
  {Couston}, {Burns}, {Pope}, {Jermyn}, {Favier}, \& {Le Bars}}]{Lecoanet2019}
{Lecoanet}, D., {Cantiello}, M., {Quataert}, E., {et~al.} 2019, \apjl, 886,
  L15, \dodoi{10.3847/2041-8213/ab5446}

\bibitem[{{Lomb}(1976)}]{Lomb1976}
{Lomb}, N.~R. 1976, \apss, 39, 447, \dodoi{10.1007/BF00648343}

\bibitem[{{MacDonald} \& {Petit}(2019)}]{MacDonald2019}
{MacDonald}, J., \& {Petit}, V. 2019, \mnras, 487, 3904,
  \dodoi{10.1093/mnras/stz1545}

\bibitem[{{Moffat} {et~al.}(2008){Moffat}, {Marchenko}, {Zhilyaev}, {Rowe},
  {Muntean}, {Chen{\'e}}, {Matthews}, {Kuschnig}, {Guenther}, {Rucinski},
  {Sasselov}, {Walker}, \& {Weiss}}]{Moffat2008}
{Moffat}, A.~F.~J., {Marchenko}, S.~V., {Zhilyaev}, B.~E., {et~al.} 2008,
  \apjl, 679, L45, \dodoi{10.1086/589237}

\bibitem[{{Paxton} {et~al.}(2011){Paxton}, {Bildsten}, {Dotter}, {Herwig},
  {Lesaffre}, \& {Timmes}}]{Paxton2011}
{Paxton}, B., {Bildsten}, L., {Dotter}, A., {et~al.} 2011, \apjs, 192, 3,
  \dodoi{10.1088/0067-0049/192/1/3}

\bibitem[{{Paxton} {et~al.}(2013){Paxton}, {Cantiello}, {Arras}, {Bildsten},
  {Brown}, {Dotter}, {Mankovich}, {Montgomery}, {Stello}, {Timmes}, \&
  {Townsend}}]{Paxton2013}
{Paxton}, B., {Cantiello}, M., {Arras}, P., {et~al.} 2013, \apjs, 208, 4,
  \dodoi{10.1088/0067-0049/208/1/4}

\bibitem[{{Paxton} {et~al.}(2015){Paxton}, {Marchant}, {Schwab}, {Bauer},
  {Bildsten}, {Cantiello}, {Dessart}, {Farmer}, {Hu}, {Langer}, {Townsend},
  {Townsley}, \& {Timmes}}]{Paxton2015}
{Paxton}, B., {Marchant}, P., {Schwab}, J., {et~al.} 2015, \apjs, 220, 15,
  \dodoi{10.1088/0067-0049/220/1/15}

\bibitem[{{Paxton} {et~al.}(2018){Paxton}, {Schwab}, {Bauer}, {Bildsten},
  {Blinnikov}, {Duffell}, {Farmer}, {Goldberg}, {Marchant}, {Sorokina},
  {Thoul}, {Townsend}, \& {Timmes}}]{Paxton2018}
{Paxton}, B., {Schwab}, J., {Bauer}, E.~B., {et~al.} 2018, \apjs, 234, 34,
  \dodoi{10.3847/1538-4365/aaa5a8}

\bibitem[{{Paxton} {et~al.}(2019){Paxton}, {Smolec}, {Schwab}, {Gautschy},
  {Bildsten}, {Cantiello}, {Dotter}, {Farmer}, {Goldberg}, {Jermyn}, {Kanbur},
  {Marchant}, {Thoul}, {Townsend}, {Wolf}, {Zhang}, \& {Timmes}}]{Paxton2019}
{Paxton}, B., {Smolec}, R., {Schwab}, J., {et~al.} 2019, \apjs, 243, 10,
  \dodoi{10.3847/1538-4365/ab2241}

\bibitem[{{Ricker} {et~al.}(2015){Ricker}, {Winn}, {Vanderspek}, {Latham},
  {Bakos}, {Bean}, {Berta-Thompson}, {Brown}, {Buchhave}, {Butler}, {Butler},
  {Chaplin}, {Charbonneau}, {Christensen-Dalsgaard}, {Clampin}, {Deming},
  {Doty}, {De Lee}, {Dressing}, {Dunham}, {Endl}, {Fressin}, {Ge}, {Henning},
  {Holman}, {Howard}, {Ida}, {Jenkins}, {Jernigan}, {Johnson}, {Kaltenegger},
  {Kawai}, {Kjeldsen}, {Laughlin}, {Levine}, {Lin}, {Lissauer}, {MacQueen},
  {Marcy}, {McCullough}, {Morton}, {Narita}, {Paegert}, {Palle}, {Pepe},
  {Pepper}, {Quirrenbach}, {Rinehart}, {Sasselov}, {Sato}, {Seager},
  {Sozzetti}, {Stassun}, {Sullivan}, {Szentgyorgyi}, {Torres}, {Udry}, \&
  {Villasenor}}]{Ricker2015}
{Ricker}, G.~R., {Winn}, J.~N., {Vanderspek}, R., {et~al.} 2015, Journal of
  Astronomical Telescopes, Instruments, and Systems, 1, 014003,
  \dodoi{10.1117/1.JATIS.1.1.014003}

\bibitem[{{Scargle}(1982)}]{Scargle1982}
{Scargle}, J.~D. 1982, \apj, 263, 835, \dodoi{10.1086/160554}

\bibitem[{{Schultz} {et~al.}(2020){Schultz}, {Bildsten}, \&
  {Jiang}}]{Schultz2020}
{Schultz}, W.~C., {Bildsten}, L., \& {Jiang}, Y.-F. 2020, \apj, 902, 67,
  \dodoi{10.3847/1538-4357/abb405}

\bibitem[{{Sim{\'o}n-D{\'\i}az} {et~al.}(2018){Sim{\'o}n-D{\'\i}az}, {Aerts},
  {Urbaneja}, {Camacho}, {Antoci}, {Fredslund Andersen}, {Grundahl}, \&
  {Pall{\'e}}}]{Simon-Diaz2018}
{Sim{\'o}n-D{\'\i}az}, S., {Aerts}, C., {Urbaneja}, M.~A., {et~al.} 2018, \aap,
  612, A40, \dodoi{10.1051/0004-6361/201732160}

\bibitem[{{Sim{\'o}n-D{\'\i}az} {et~al.}(2017){Sim{\'o}n-D{\'\i}az}, {Godart},
  {Castro}, {Herrero}, {Aerts}, {Puls}, {Telting}, \&
  {Grassitelli}}]{Simon-Diaz2017}
{Sim{\'o}n-D{\'\i}az}, S., {Godart}, M., {Castro}, N., {et~al.} 2017, \aap,
  597, A22, \dodoi{10.1051/0004-6361/201628541}

\bibitem[{{Sim{\'o}n-D{\'\i}az} {et~al.}(2014){Sim{\'o}n-D{\'\i}az}, {Herrero},
  {Sab{\'\i}n-Sanjuli{\'a}n}, {Najarro}, {Garcia}, {Puls}, {Castro}, \&
  {Evans}}]{SimonDiaz2014}
{Sim{\'o}n-D{\'\i}az}, S., {Herrero}, A., {Sab{\'\i}n-Sanjuli{\'a}n}, C.,
  {et~al.} 2014, \aap, 570, L6, \dodoi{10.1051/0004-6361/201424742}

\bibitem[{{Sim{\'o}n-D{\'\i}az} {et~al.}(2010){Sim{\'o}n-D{\'\i}az}, {Herrero},
  {Uytterhoeven}, {Castro}, {Aerts}, \& {Puls}}]{Simon-Diaz2010}
{Sim{\'o}n-D{\'\i}az}, S., {Herrero}, A., {Uytterhoeven}, K., {et~al.} 2010,
  \apjl, 720, L174, \dodoi{10.1088/2041-8205/720/2/L174}

\bibitem[{{Stone} {et~al.}(2020){Stone}, {Tomida}, {White}, \&
  {Felker}}]{Stone2020}
{Stone}, J.~M., {Tomida}, K., {White}, C.~J., \& {Felker}, K.~G. 2020, \apjs,
  249, 4, \dodoi{10.3847/1538-4365/ab929b}

\bibitem[{{Sundqvist} {et~al.}(2013){Sundqvist}, {Petit}, {Owocki}, {Wade},
  {Puls}, \& {MiMeS Collaboration}}]{Sundqvist2013}
{Sundqvist}, J.~O., {Petit}, V., {Owocki}, S.~P., {et~al.} 2013, \mnras, 433,
  2497, \dodoi{10.1093/mnras/stt921}

\bibitem[{{Townsend}(2010)}]{Townsend2010}
{Townsend}, R.~H.~D. 2010, \apjs, 191, 247, \dodoi{10.1088/0067-0049/191/2/247}

\bibitem[{Virtanen {et~al.}(2020)Virtanen, Gommers, Oliphant, Haberland, Reddy,
  Cournapeau, Burovski, Peterson, Weckesser, Bright, {van der Walt}, Brett,
  Wilson, Millman, Mayorov, Nelson, Jones, Kern, Larson, Carey, Polat, Feng,
  Moore, {VanderPlas}, Laxalde, Perktold, Cimrman, Henriksen, Quintero, Harris,
  Archibald, Ribeiro, Pedregosa, {van Mulbregt}, \& {SciPy 1.0
  Contributors}}]{Scipy2020}
Virtanen, P., Gommers, R., Oliphant, T.~E., {et~al.} 2020, Nature Methods, 17,
  261, \dodoi{10.1038/s41592-019-0686-2}

\end{thebibliography}
\bibliographystyle{aasjournal}


\end{CJK*}
\end{document}